# Robust Topological Bound States in the Continuum in a Quantum Hall Bar with an Anti-dot

Ricardo Y. Díaz and Carlos Ramírez*

Departamento de Física, Facultad de Ciencias, Universidad Nacional Autónoma de México, Apartado Postal 70542, Ciudad de México 04510, México

*Corresponding author e-mail address: carlos@ciencias.unam.mx

**Abstract**

Bound states in the continuum (BICs) are quantum states with normalizable wave functions and energies that lie within the continuous spectrum for which extended or dispersive states are also available. These special states, which have shown great applicability in photonic systems for devices such as lasers and sensors, are also predicted to exist in electronic low-dimensional solid-state systems. The non-trivial topology of materials is within the known mechanisms that prevent the bound states to couple with the extended states. In this work we search for topologically protected BICs in a quantum Hall bar with an anti-dot formed by a pore far from the borders of the bar. The bound state energies and wavefunctions are calculated by means of the Recursive S-Matrix method. The resulting bound state energies coexist with extended states and exhibit a pattern complimentary to the Hofstadter butterfly. A symmetry-breaking diagonal disorder was introduced, showing that the BICs with energies far from the Landau levels remain robust. Moreover, the energy difference between consecutive BICs multiplied by the anti-dot perimeter follows the same curve despite disorder. Finally, a BIC-mediated current switching effect was found in a multi-terminal setup, which might permit their experimental detection.

**Introduction**

In 1929, Wigner and von Neumann introduced a potential in the Schrödinger equation that leads to quantum states with normalizable wavefunctions that coexist with the continuum energy spectrum of extended states [1]. These special states are called Bound States in the Continuum (BICs) and for many years they were considered a mathematical curiosity. In recent years, BICs emerged as a valuable area of research due to their appearance, both theoretically and in the experiment, in photonic [2–12], acoustic [13–17], mechanical [18,19], electronic [20–25] and anyonic [26] systems. Since their observation, BICs have been proposed as the working principle of devices such as lasers [27–30], transductors [19] and filters [31]. They have also been proposed for qubit encryption [32], as well as for optical [33–35], thermal [36] and biological [37] sensors. Electronic BICs have been predicted in low-dimensional solid-state systems such as quantum-dot arrays [20–22], topological materials [23,24], disordered structures [17] and one-dimensional interacting systems [38]. The experimental detection of electronic BICs has been aided by setups made of photonic crystals [39] and electronic circuits [26,40], which can be designed to have a tight-binding like description.

From a theoretical perspective, BIC identification is challenging. Extended states in the continuum in one-dimensional nanostructures require infinite or semi-infinite periodic leads. However, this periodicity must be deliberately broken for bound states to appear, discarding the use of Bloch's theorem. On the other hand, direct diagonalization is impractical as the Hamiltonian matrix is of infinite size. Considering a truncated version of the Hamiltonian, BICs can be approximately found by calculating the inverse participation ratio (IPR) for every wavefunction [17]. Alternatively, the eigenenergies can be compared with the ones calculated for a bigger portion of the system; the eigenvalues that do not depend on the size of the truncated Hamiltonian might be associated to localized wavefunctions and can be considered potential BICs [38]. Another method consists in modeling open systems with an effective non-Hermitian Hamiltonian so that the real eigenvalues can be associated to localized wavefunctions [23]. As BICs are not expected to contribute to conductance, a simple way of finding them is to search the energies for which the density of states (DOS) has a peak that does not appear in the conductance [21]. BICs can also be found by introducing the effect of infinite leads in a matrix equation as self-energy sub-matrices and looking for the zero singular values as a function of the energy [41]. A technique based on the Recursive S-Matrix Method (RSMM) has shown to be able to determine BIC energies and wavefunctions with great precision and low computational cost due to its modular implementation and optimization by a divide and conquer algorithm [20,42].

For a BIC to exist, the loss of periodicity is necessary but not sufficient as there must be a mechanism that prevents the bound state to couple with the extended states. Symmetry, separability of the Hamiltonian, interference, and topological protection are among the BIC protection mechanisms known to date [2,23,24]. Here we focus on a kind of topologically protected BICs.

The topological nature of an insulator is determined by a quantity called *topological invariant* that remains unaltered unless de band gap is closed. An insulator with a topological invariant different from that of the vacuum is said to be in a non-trivial topological phase and is referred to as a topological insulator (TI). A well-known feature of TIs is the appearance of edge states localized in the border of the sample with energies within the bulk band gap. Topological edge states depend exclusively on the non-trivial topology of the material, which makes them robust against defects such as disorder or border irregularities. Therefore, TIs are often considered great candidates for achieving dissipationless electronic transport via edge states [43] or for quantum information transfer [44,45] despite disorder.

A well-known scenario where multiple topological phases arise is the integer quantum Hall effect (QHE) that occurs in a 2-dimensional (2D) material subjected to a non-coplanar magnetic field. If the applied field is high enough a set of chiral edge states (CES) appear in the border of the material, producing a quantized resistance $R_{XY} = h/\nu e^2$ transverse to the current flow, where $e$ is the electron charge, $h$ is the Planck constant and $\nu$ is an integer filling factor that changes discretely for energies known as Landau levels [46]. The connection with topology is given as the filling factor $\nu$ is equal to a topological invariant known as the TKNN invariant or Chern number $C$ [47]. An interesting question is under which conditions the CES can remain decoupled from the extended states. Whenever this happens, topological BICs may appear with properties inherited from their topological nature such as their protection against disorder. As the synthesis of nanostructures is almost inevitably affected by defects, the determination of BIC energies robust against disorder might be crucial to achieve their experimental observation.

Since the original experiment made by von Klitzing [48], multiple variants of the QHE have been introduced to further explore topological states in quantum materials. Such is the case of the fractional QHE in graphene [49] and the quantum anomalous Hall effect in intrinsic TIs [50–52]. The high tunability and robustness of the QHE are desirable features to observe electronic BICs by their effects in measurable quantities. For instance, non-zero longitudinal resistance in a quantum Hall bar with a periodic distribution of anti-dots (pores) was suggested as consequence of states localized in the anti-dots [53]. Moreover, recently proposed techniques have shown promising results for the synthesis of effective 2D structures with tailor made features such as pores [54,55] and multiple magnetic dominions with arbitrary boundary forms [51] or arbitrary Chern number difference between adjacent dominions [50,52]. Therefore, quantum Hall bars with pores or multiple magnetic dominions are ideal candidates for exploring the relation between bound states and measurable quantities such as magnetoresistance or conductance.

In this work, we search topological BICs in a quantum Hall bar with an anti-dot (pore) in the center of the bar, as shown in Fig. 1a. We introduce a continuous of extended states by connecting infinite periodic leads to the Hall bar. The applied magnetic field induces chiral edge states (CES) in the outer border of the sample that couple with the leads producing a finite conductance. The CES are also expected to appear in the anti-dot border. As the anti-dot CES decay exponentially into the bulk of the bar, we expect them to be decoupled from the extended states provided by the leads. This might allow the anti-dot border CES to be identified as BICs. The primary objective of this paper is to confirm the existence of these BICs, and analyze their energy spectra, wavefunctions, and their influence on electronic transport properties. In section 2 we show the bound state energies and wave functions obtained for this system. In section 3 a diagonal disorder is applied to examine the robustness of these BICs. Finally, the relation between the BIC energies and the currents in a multi-terminal quantum Hall bar is presented and discussed in section 4.

## 2. BIC energy spectrum and wave functions

Let us consider a 2D square lattice of $N \times M$ sites with an anti-dot formed by a pore that removes $N_A \times M_A$ sites in the center of the sample, as shown in Fig. 1a. A uniform magnetic field is applied perpendicular to the surface which produces a non-zero magnetic flux $\Phi$ per unit cell. The right and left sides of the sample are connected to leads formed by $N_L$ infinite and periodic chains. The system is described by the tight-binding Hamiltonian

$$\hat{H} = \sum_n \left( \varepsilon_0 \hat{c}_n^\dagger \hat{c}_n + \sum_{n \neq m} t_{n,m} \hat{c}_m^\dagger \hat{c}_n \right), \tag{1}$$

where $\hat{c}_n^\dagger (\hat{c}_n)$ is the fermionic creation(annihilation) operator in the $n-th$ site of the lattice and the sums develop over every site in the Hall bar excluding the ones eliminated by the anti-dot. The hopping parameters $t_{n,m}$ are chosen to vanish whenever $n$ and $m$ are not nearest neighbors, otherwise they take the value $t_{n,m} = t_0 e^{i\theta_{n,m}}$ where $\theta_{n,m}$ is the Peierls phase given as [56]

$$\theta_{n,m} = \frac{e}{\hbar} \int_{\mathbf{r}_n}^{\mathbf{r}_m} \mathbf{A} \cdot d\mathbf{l}. \tag{2}$$

The term $\mathbf{r}_n$ is the position of the $n-th$ lattice site and $\mathbf{A}$ is the magnetic vector potential. We use the Landau gauge $\mathbf{A} = \left(0, x\Phi/a^2, 0\right)$ where $a^2$ is the unit cell area. The on-site energy $\varepsilon_0$ takes the value $\varepsilon_0 = 4|t_0|$, which comes from the discretization of a continuous system [56]. The parameters chosen for the following numerical calculations are $N = 101$, $M = 100$, $N_A = 45$, $M_A = 44$ and $N_L = 30$. BIC determination is made using the technique based on the Recursive S-Matrix Method (RSMM) [20], whose implementation is described in the Appendix.

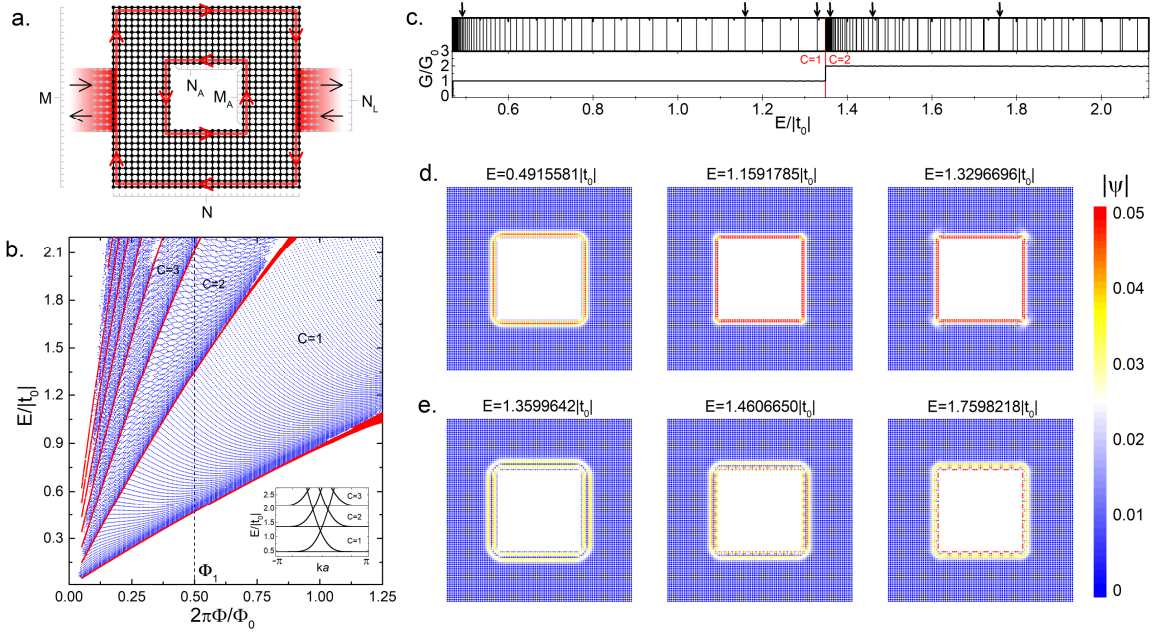

**Figure 1: a.** Representation of the square-lattice quantum Hall bar with an anti-dot and infinite periodic leads. Red lines represent schematically the topological CES in the system. The CES in the outer border interact with incoming and outgoing waves in the leads. For $N=100$, $M=100$, $N_A=45$, $M_A=44$ and $N_L=30$. **b.** BIC energy spectrum (blue dots) of the system as a function of magnetic flux per unit cell $\Phi$. A portion of the Hofstadter butterfly is shown in red. **Inset:** the band structure of an $M=100$ sites tall infinite nanoribbon and a magnetic flux $\Phi_1 = (1/4\pi)\Phi_0$. **c.** BIC energy spectrum (upper panel) and conductance between the left and right leads (lower panel). **d.** and **e.** BIC wave function amplitudes corresponding to $C=1$ and $C=2$, respectively. The magnetic flux considered in c., d. and e. is also $\Phi_1$.

In Fig. 1b, we present the bound state energy spectrum (blue dots) as a function of $\Phi/\Phi_0$, where $\Phi_0 = 2\pi\hbar/e$ is the magnetic flux quantum. This figure also shows in red color a portion of the Hofstadter butterfly that identifies the changes in the Chern number [57]. Notice that the bound state energy spectrum is discrete when $\Phi$ is held constant. However, as $\Phi$ changes, the energy spectrum evolves smoothly. It can also be seen that the density of the spectrum becomes larger near the energies where the Chern number increases. For instance, for $C=1$ the separation between consecutive bound state energies grows monotonically, while for $C>1$ a more complex spectrum arises due to the addition of new CES.

The wavefunction of the outer CES couple with the lead modes, forming extended states with energies in the continuum. The band structure associated to the outer CES is show in the inset. This was made for an infinite nanoribbon of width $M=100$ for a magnetic flux $\Phi_1 = \frac{1}{4\pi}\Phi_0$, which is marked with a dashed line in Fig. 1b. The Chern number can be identified in the inset as it is related to the number of bands present for a given energy. Hence, extended states will always be present as long as $C>0$. This observation is reaffirmed by the lower panel of Fig. 1c that shows the left-to-right Landauer conductance [58]. Notice that the conductance appears in multiples of the conductance quantum $G_0 = 2e^2/\hbar$, where the integer multiple is equal to the Chern number. This proves that the conductance is mediated by outer CES that appear for a continuous energy spectrum. The upper panel of the Fig. 1c presents the bound state energies (black lines) for the same magnetic flux. Notice that for all these energies we have $C>0$, confirming that they are indeed BIC energies.

We select the energies indicated with arrows in Fig. 1c and calculate their wavefunctions, which are shown in Figs 1d and 1e for $C=1$ and $C=2$, respectively. In all cases presented, the wave function remains localized near the anti-dot border, which is consistent with the expected exponential decay of topological CES into the bulk. This rapid decay far from the anti-dot border is what prevents the BICs from coupling with the extended states that lie close to the outer border. Observe how the $C=2$ wavefunctions in Fig. 1e present nodes and penetrate further into the bulk compared to the $C=1$ wavefunctions of Fig. 1d, where single CES are allowed. This occurs because multiple CES are allowed for $C>1$.

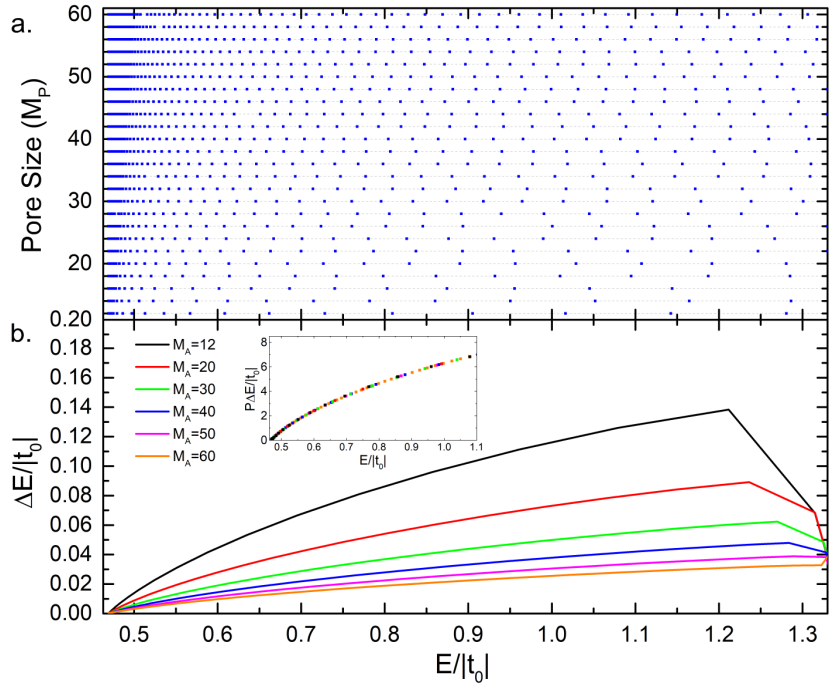

**Figure 2: a.** BIC energies and **b.** difference $\Delta E$ between consecutive BIC energies (lower panel) for anti-dots of size $M_A \times (M_A + 1)$. In all cases, the anti-dot was embedded in a Hall bar as the one shown in Fig. 1a with $N=100$, $M=100$ and $N_L = 30$. The inset shows $\Delta E$ multiplied by the anti-dot border perimeter $P$ for each anti-dot size.

The size of the anti-dot determines the density of the discrete spectrum. This fact can be observed in Fig. 2a where the BIC energies with $C=1$ are presented for different anti-dot sizes. As the anti-dot becomes larger, the density of the BIC energy spectrum grows as well. Moreover, we present in Fig. 2b the difference $\Delta E$ between consecutive BIC energies for different anti-dot sizes. A monotonically increasing behavior can be observed for all cases with a slope that depends on the anti-dot size. Observe the inset in the Fig. 2b, where we plot $\Delta E$ multiplied by the anti-dot border perimeter $P$, and notice how $P\Delta E$ falls into the same curve independently of the anti-dot size. This shows that the separation between BIC energies behaves equivalently for every anti-dot size up to a constant multiplying factor.

## 3. Robustness against disorder

Topological edge states are protected by their topological invariant value, so the BICs presented in section 2 are expected to remain robust over perturbations that do not change the Chern number. To determine whether this robustness is present or not, we consider a diagonal disorder in Hamiltonian (1) by adding the term

$$\hat{H}_{dis} = \sum_n \varepsilon_n \hat{c}_n^\dagger \hat{c}_n. \tag{3}$$

where $\varepsilon_n$ is an additional on-site energy term that varies from site to site. To control the amount of disorder let us define a disorder concentration $\rho \in [0,1]$ that determines the portion of randomly selected sites that will have a non-zero value of $\varepsilon_n$, this can be thought as having impurities in the Hall bar. The impurity on-site energies follow a uniform random distribution from the interval $[-W/2, W/2]$ where $W$ is the disorder strength. This proposed form of disorder breaks the spatial symmetries even for low disorder concentrations and strengths.

We present in Fig. 3a the BIC energies calculated for multiple realizations with disorder concentrations of $\rho = 0.05$, $\rho = 0.15$ and $\rho = 0.5$, considering in all cases a disorder strength of $W = |t_0|$. Observe that the BIC energies near the Landau levels in the pristine bar are no longer identified as BIC energies as the disorder concentration increases. The Landau levels are known for having a high DOS associated to bulk states, which are responsible for the Shubnikov-de Haas oscillations in the Hall resistance [53]. The disorder term (3) provokes a widening of the high DOS energy interval near the Landau levels [59]. This widening allows the CES to couple with the bulk states destroying the topological isolation between inner and outer CES, which explains the absence of BICs around Landau levels. However, it can also be observed that far from the Landau levels the BICs are still present. We calculated the wave functions of the surviving BICs for different values of $\rho$ and show them in Fig. 3b. Notice that even though there are slight variations with respect to the wave functions presented in figures 1d and 1e, the surviving BICs are still confined to the anti-dot border and continue to have a behavior dependent of the Chern number. As they persist despite disorder and preserve their amount of localization, we will be referring to the surviving BICs far from the Landau levels as robust BICs.

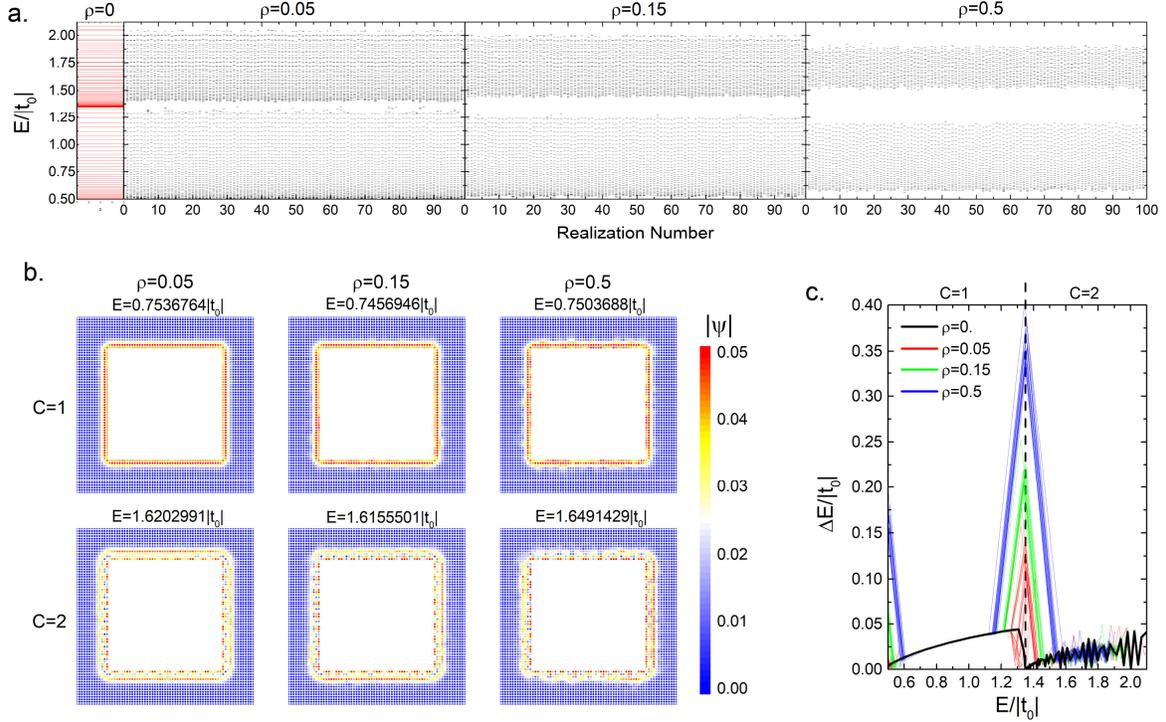

**Figure 3**: **a.** BIC energy spectrum for different disorder concentrations $\rho$ and a disorder strength of $W = |t_0|$. **b.** Example of robust BIC wave functions. **c.** Difference between consecutive BIC energies for different values of $\rho$.

It can be observed from Fig. 3a that the robust BIC energies vary between realizations even for the same disorder concentration. However, we can still ask ourselves if the separation between consecutive BIC energies remains invariant. We present in Fig. 3c the separation $\Delta E$ between consecutive BIC energies for multiple realizations and different values of $\rho$. Notice how in the $C = 1$ region there is an interval in which all of the curves are identical even though they correspond to different disorder concentrations. This result shows that the amount of disorder determines the energy interval in which robust BICs can exist, but the separation between consecutive BICs within this region remains invariant and exhibits the same behavior as in the case without disorder of Fig. 2b.

## 4. Disorder-Assisted Effects in quantum Hall currents

For the BICs presented in the previous sections to be experimentally detected it is necessary to identify a measurable physical quantity associated with them. For this end, we propose the well-known six-terminal quantum Hall bar as the one presented in Fig. 4a. We model each terminal as formed by $N_L$ periodic chains attached to the system. To identify BICs by measuring transport properties there must be a coupling between them and the outer CES that contribute to the conductance. Thus, the anti-dot must be close to the border of the bar for the BICs to interact with the outer CES. In the following, we will consider a Hall bar of $N = 161$ sites long per $M = 52$ sites tall with an antidot of $N_A = 45$ sites long per $M_A = 44$ sites tall in the center of the bar, which is the same size of the anti-dot whose BIC energies were presented in sections 2 and 3. There are $N_s = 70$ sites separating the terminals 2 and 3 as well as terminals 5 and 6.

At zero temperature, the current $I_n$ flowing through the $n-th$ terminal can be calculated by the Landauer-Büttiker formula [53,58]

$$I_n = \frac{2e}{h} \sum_m \left( T_{m,n}(E) V_n - T_{n,m}(E) V_m \right), \quad (4)$$

where $V_n$ represents the voltage in the $n-th$ terminal and $T_{n,m}$ is the transmission coefficient from the $n-th$ terminal into the $m-th$ terminal. Let us consider the case where a bias potential $V_S$ is established in terminal 1, while terminals 3, 4 and 6 are grounded. Terminals 2 and 5 will be considered as floating so that the net current in those terminals vanishes. This configuration allows to explore BIC-related effects in how the current is distributed between the grounded terminals. The values of $T_{n,m}(E)$ are obtained directly from the S-matrices calculated by the RSMM [60].

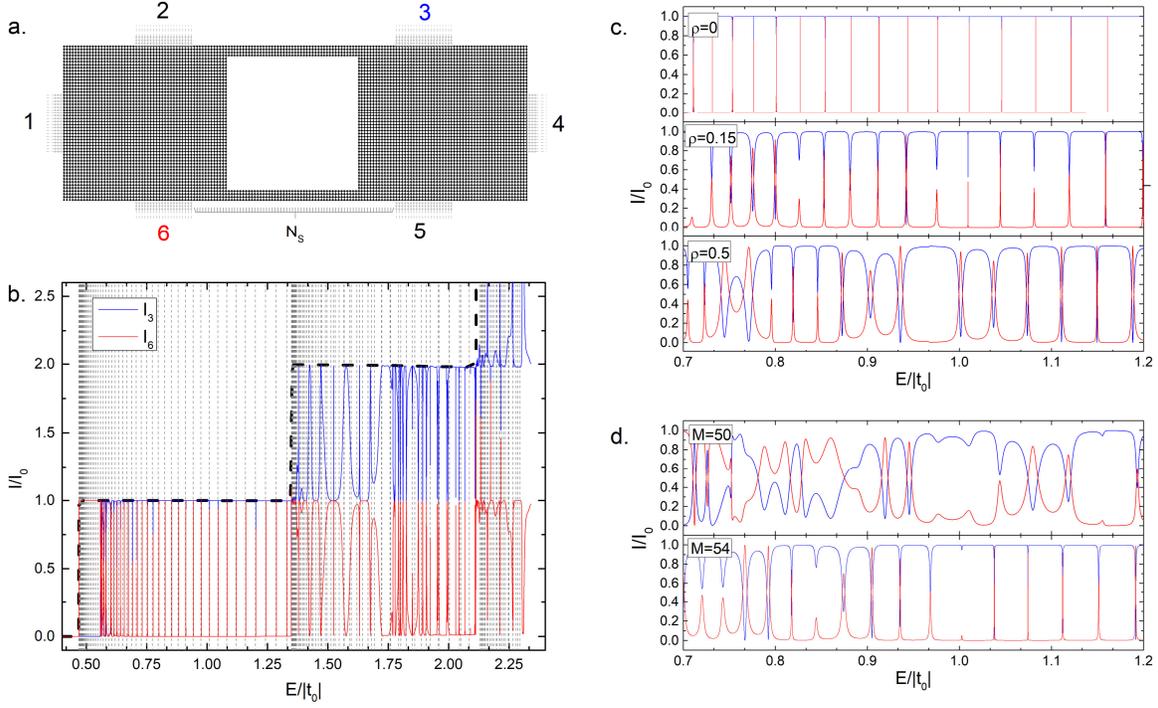

**Figure 4: a.** Schematic representation of a quantum Hall bar with 6 terminals. **b-d.** Currents $I_3$ and $I_6$ flowing through terminals 3 (blue) and 6 (red) respectively as a function of the energy for a magnetic flux $\Phi_1$. The thin dashed lines in **b.** correspond to BIC energies while the thick dashed line is the current $I_3$ when there is no anti-dot. In **c.** the bar dimensions are held constant while the disorder concentration $\rho$ varies. The currents in **d.** are calculated for a fixed disorder concentration $\rho = 0.5$ and different bar widths $M$.

Figure 4b shows the results obtained from equation (4) for the currents $I_3$ and $I_6$ that flow through the terminals 3 and 6, respectively. The currents are presented in terms of a reference current $I_0$, which is the total current running through terminal 1 when a single CES is present, that is, for $C=1$. Thin dashed lines indicate the BIC energies of the anti-dot. The BIC energies match with sharp switches between the currents $I_3$ and $I_6$. In contrast, for the bar without anti-dot (thick dashed line) the current flows entirely through terminal 3 with $I_3 = CI_0$. This suggests that the sharp switches between the currents are caused by BICs in the edge of the anti-dot which produce pathways to terminal 6.

If the energy interval for which the currents switch is too small, the effect might be hard to observe in the experiment because of insufficient instrument precision or due to temperature. Thus, it is desirable to reduce the sharpness of the current oscillations. The switches between $I_3$ and $I_6$ depend on the interaction between CES in the border of the bar and the anti-dot, so the aim is to enhance the interaction between two exponentially decaying CES. This can be achieved by adding disorder or by having the CES closer to each other [50]. First, let us consider a Hall bar and anti-dot of the same sizes as the ones used to obtain Fig. 4b, but including a diagonal disorder as the one defined in Section 3. The currents calculated under these conditions for $W = |t_0|$ and different values of $\rho$ are presented in Fig. 4c. Notice that as $\rho$ increases the energy intervals for which the currents switch get wider. An important remark is that even though the robust BIC energies get displaced by the disorder, the energy separation $\Delta E$ between consecutive states remains unaltered and follows the pattern shown in Fig. 3c. Likewise, the energy separation between adjacent current peaks remains invariant in presence of disorder. Therefore, the presence of disorder in the Hall bar is desirable to an extent for the BIC-mediated current switch to be observed in the experiment.

Let us now consider the role of proximity between the anti-dot and the border of the bar. Figure 4d shows the currents calculated for a fixed disorder concentration of $\rho = 0.5$ and $W = |t_0|$, changing the Hall bar width from the original value of $M = 52$ to $M = 50$ and $M = 54$. Observe how for $M = 50$, where the anti-dot is closer to the border of the bar, the oscillations are still present but the overall behavior of both currents becomes more distorted with respect to the one observed when the anti-dot is further from the border. On the other hand, the oscillations recover their sharpness for $M = 54$. This analysis shows that both the disorder in the bar and the proximity between the anti-dot and the border determine the width of the energy intervals for which the currents switch.

## 5. Conclusions

In this work we determined the presence of BICs in a quantum Hall bar with an anti-dot formed by a rectangular-shaped pore. The BIC energies developed a pattern that compliments a Hofstadter butterfly where the density of the spectrum depends on the anti-dot size. We observed that the separation between consecutive BIC energies for Chern number $C = 1$ multiplied by the perimeter of the anti-dot, follows the same pattern for every anti-dot size. All the calculated BIC wavefunctions were localized around the anti-dot border and it was found that the $C = 2$ states penetrate further into the bulk than the $C = 1$ ones. A diagonal disorder was introduced, showing that the BICs with energies far from the Landau levels persist under such conditions. Even though the robust BIC energies are displaced by disorder, the difference between consecutive robust BIC energies is the same as in the case where there is no disorder. Finally, a six-terminal quantum Hall bar was analyzed, finding a current switching effect in the BIC energies. It was also observed that the sharpness of the current switching as a function of energy can be controlled by varying the amount of disorder in the bar as well as the separation between the anti-dot and the bar border.

The precision and diversity of the techniques with which pores can be introduced in effective 2D structures have grown in recent years [54,55], so the BIC-mediated current switching effect represents an opportunity for the experimental detection of electronic BICs in low-dimensional solids. Moreover, we do not expect this effect to be restricted to the single anti-dot case. The effect can be further explored by considering multiple anti-dots o even a complex anti-dot lattice, as was made in previous magnetoresistance studies [53,61]. Another option is to replace the anti-dots with different magnetic domains. For example, a quantum Hall bar that instead of an anti-dot has a magnetic domain with a Chern number different from that of the rest of the bar. This last setup would be closely related to the Quantum-Anomalous-Hall-Effect that occurs in intrinsic TIs [50–52]. It is within the scope of the authors' future research to consider these cases to study BIC-related physical phenomena.

## Acknowledgments

This work has been supported by UNAM-DGAPA-PAPIIT IN109022. Computations were performed at Miztli under project LANCAD-UNAM-DGTIC-329. Ricardo Y. Díaz thanks CONAHCyT for the schorlarship granted.

## Appendix: BIC determination method

We calculate the BIC energies and wavefunctions assisted by the RSMM following the method described in Ref. [20]. This method requires to define two systems, A and B, so that their combination through the RSMM results in the system for which BICs are to be determined. We define system A by adding $N_{Aux}$ auxiliary chains to the red sites of Fig. 5a following a vertical line below the anti-dot. On the other hand, system B is formed by $N_{Aux}$ independent sites, each of them with an auxiliary chain as shown in Fig. 5b. Auxiliary chains are one-dimensional atomic chains with null on-site energy and a constant hopping integral $t_C > E/2$. Incoming and outgoing waves travel through auxiliary chains. To merge systems A and B by the RSMM, the wave amplitudes incoming from the auxiliary chains in system A are made equal to the outgoing waves in system B, and vice versa. This results in the system of Fig. 1a.

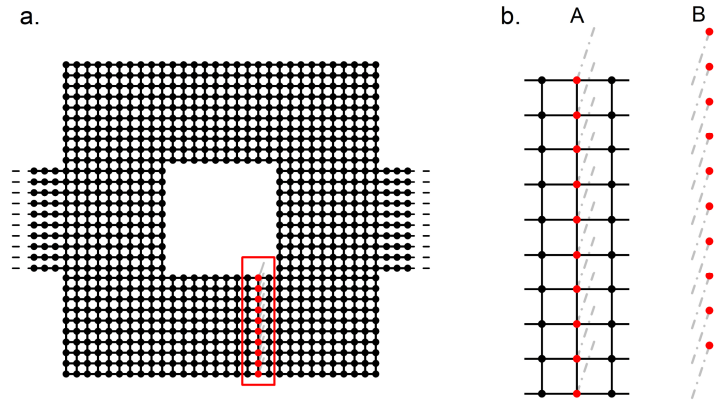

**Figure 5: a.** Schematic illustration of system A, formed by a Hall bar with $N_{Aux}$ auxiliary chains (represented by gray lines) added to the sites marked in red. **b.** Zoom to the auxiliary chains attached in system A and complementary system B formed by $N_{Aux}$ independent sites with auxiliary chains attached.

The complete S-matrix of the system shown in Fig. 5a can be written in the form

$$\mathbf{S} = \begin{pmatrix} \mathbf{S}_{AA} & \mathbf{S}_{AL} \\ \mathbf{S}_{LA} & \mathbf{S}_{LL} \end{pmatrix}, \quad (5)$$

where $A$ denotes the auxiliary chains, $L$ denotes the leads, and $\mathbf{S}_{\sigma'\sigma}$ describes the scattering from waves incoming from $\sigma$ and dispersed to $\sigma'$. On the other hand, an S-matrix $\mathbf{S}_B$ can be calculated to describe the scattering from the auxiliary chains in system B into themselves. According to Ref. [20], bound state energies are the ones for which

$$\mathbf{B}^+ = \mathbf{S}_{AA}\mathbf{S}_B\mathbf{B}^+, \quad (6)$$

*i.e.*, whenever $\mathbf{S}_{AA}\mathbf{S}_B$ has an eigenvalue equal to one. We use a root finding algorithm (secant) to numerically determine the energies for which the imaginary this happens. It is also shown in Ref. [20], that the wave function coefficients of sites with attached auxiliary chains can be calculated in terms of the eigenvector of equation (6) as $\mathbf{C} = (\mathbf{I} + \mathbf{S}_B)\mathbf{B}^+$ where $\mathbf{I}$ is the identity matrix and $\mathbf{C}$ is a vector containing the wavefunction coefficients. By varying the position and number of the auxiliary chains, the wavefunction in the complete Hall bar can be calculated.